\documentclass[a4paper]{article}

\usepackage{INTERSPEECH2021}
\usepackage{multirow}
\usepackage{cite}
\usepackage{hyperref}
\makeatletter
\newcommand{\thickhline}{%
	\noalign {\ifnum 0=`}\fi \hrule height 1.0pt
	\futurelet \reserved@a \@xhline
}

\title{Learning Explicit Prosody Models and Deep Speaker Embeddings for \\ Atypical Voice Conversion}
\name{Disong Wang$^1$, Songxiang Liu$^1$, Lifa Sun$^2$, Xixin Wu$^3$, Xunying Liu$^1$, Helen Meng$^1$}
\address{$^1$Human-Computer Communications Laboratory\\The Chinese University of Hong Kong, Hong Kong SAR, China\\$^2$SpeechX Limited, Shenzhen, China\\$^3$Department of Engineering, University of Cambridge, UK}
\email{\{dswang, sxliu, xyliu, hmmeng\}@se.cuhk.edu.hk, lfsun@speechx.cn, xw369@cam.ac.uk}

\begin{document}

\maketitle
\begin{abstract}
  Though significant progress has been made for the voice conversion (VC) of typical speech, VC for atypical speech, e.g., dysarthric and second-language (L2) speech, remains a challenge, since it involves correcting for atypical prosody while maintaining speaker identity. To address this issue, we propose a VC system with explicit prosodic modelling and deep speaker embedding (DSE) learning. First, a speech-encoder strives to extract robust phoneme embeddings from atypical speech. Second, a prosody corrector takes in phoneme embeddings to infer typical phoneme duration and pitch values. Third, a conversion model takes phoneme embeddings and typical prosody features as inputs to generate the converted speech, conditioned on the target DSE that is learned via speaker encoder or speaker adaptation. Extensive experiments demonstrate that speaker adaptation can achieve higher speaker similarity, and the speaker encoder based conversion model can greatly reduce dysarthric and non-native pronunciation patterns with improved speech intelligibility. A comparison of speech recognition results between the original dysarthric speech and converted speech show that absolute reduction of 47.6\% character error rate (CER) and 29.3\% word error rate (WER) can be achieved. 
\end{abstract}

\section{Introduction}

Voice conversion (VC) is a technique for converting non-linguistic and para-linguistic information, such as speaker identity \cite{mohammadi2017overview}, prosody \cite{rentzos2003transformation} and accent \cite{oyamada2017non}, with potential applications in assistive speech technologies and language acquisition technologies \cite{nakamura2012speaking,felps2009foreign}. This work aims to apply VC techniques to convert atypical speech to a typical form. Specifically, we consider two types of atypical speech \cite{shor2019personalizing} – dysarthric speech and second-language (L2) speech. Dysarthric speech results from neuromotor disorders \cite{kain2007improving} that cause disturbances in muscular control during articulation. L2 speech is spoken by L2 learners with non-native accents \cite{scales2006language}. Both dysarthric and L2 speech exhibits atypical prosody, imprecise articulation and reduced intelligibility. These may engender substantial communication difficulties for dysarthric patients and hinder the pronunciation clarity of L2 learners. 

To enhance the quality of the atypical speech, our previous work \cite{wang2020end,liu2020end} presented an end-to-end VC (E2E-VC) method, where a speech-encoder is used to extract linguistic representations, e.g., phoneme embeddings, from the atypical speech, and a text-to-speech (TTS) decoder with attention maps phoneme embeddings to typical speech features. The speaker identity of the converted speech is controlled by the target speaker embedding produced by a speaker encoder \cite{liu2020end}. Though high-fidelity speech can be generated, the prosody, speaker similarity and speech intelligibility require further improvement.

In this paper, we propose an improved VC system, where the previous TTS-decoder with attention is broken into a prosody corrector and a conversion model. The prosody corrector contains phoneme duration and pitch predictors that are introduced to explicitly model the prosody for predicting typical phoneme duration and pitch features. The conversion model maps pitch and phoneme embeddings expanded by the duration to mel-spectrograms, conditioned on the target deep speaker embedding (DSE). To obtain effective DSE that captures speaker characteristics, two different approaches are investigated in our work: (1) Speaker encoder, where a speaker classifier trained independently is adopted to extract DSE from the reference target speech; (2) Speaker adaptation, where the DSE is jointly learned and fine-tuned with a pre-trained multi-speaker conversion model by using the target speech. We assume that the DSE obtained using the two approaches contains no prosody cues, so prosody and speaker identity are controlled by individual conditions, i.e., the prosody is controlled by phoneme duration and pitch, and speaker identity is controlled by the DSE. As a result, with the predicted typical prosody features and the effective DSE as conditions, the converted speech has typical pronunciation patterns with high speaker similarity and improved speech intelligibility.

The main advantages of the proposed approach include: (1) Explicit prosody correction to reduce dysarthric or non-native pronunciation patterns; (2) Improvements over previous methods \cite{wang2020end,liu2020end} in generating speech with enhanced speaker similarity, naturalness and intelligibility; (3) Potential extensibility to other atypical voice conversion and enhancement tasks. 

\section{Related work}

Dysarthric speech reconstruction (DSR) aims to convert dysarthric speech to be near-normal speech with higher intelligibility and naturalness. Various VC techniques have been applied for DSR. Rule-based VC modifies the temporal or frequency characteristics of speech according to specific rules \cite{rudzicz2011acoustic}. Statistical VC builds a mapping function between the acoustic features of dysarthric and normal speech \cite{wang2020end,aihara2013individuality,aihara2017phoneme,chen2020enhancing}. Significant progress has been achieved, but the converted speech has low speaker similarity. 

Accent conversion (AC) aims to convert the non-native L2 accented speech to become near-native speech. \cite{aryal2014can} proposed a GMM based VC by using vocal tract length normalization and linguistic content similarity matching. \cite{zhao2018accent,zhao2019using} utilized phonetic posteriorgrams (PPG) of the native speaker to generate target acoustic features. Although the non-native accent can be reduced, these methods require native reference utterances that may not be readily available. E2E-VC \cite{liu2020end} can effectively solve this issue, but speaker similarity needs to be improved as well.

We intend to convert dysarthric and L2 speech respectively to become near-normal and near-native speech with typical prosody, high speaker similarity and improved intelligibility. We reference multi-speaker TTS \cite{gibiansky2017deep,arik2018neural} that uses prosody features for speech synthesis, and DSE obtained via speaker encoder or speaker adaptation to control the speaker identity. Inspired by Deep Voice 2 \cite{gibiansky2017deep}, we introduce predictors of phoneme duration and pitch to attain typical values in order to generate the speech with typical (i.e., normal or native) prosody characteristics.

\section{Baseline method}

In this paper, we adopt the previously proposed E2E-VC for DSR \cite{wang2020end}  and AC \cite{liu2020end} as the baseline method. E2E-VC is composed of three components: (1) A sequence-to-sequence (seq2seq) based TTS model, e.g., Tacotron \cite{wang2017tacotron}, is first trained with transcribed \textit{typical} speech. The TTS-decoder with attention implicitly models the prosody, e.g., phoneme duration and pitch, which are inflexible to control during inference.  (2) Given the transcribed \textit{atypical} speech, a speech-encoder is trained to produce similar linguistic representations with those produced by the TTS-encoder. (3) By concatenating the speech-encoder and TTS-decoder with attention, an E2E-VC is formed to convert atypical speech to its typical version. Note that speaker similarity issue was not considered in the DSR work \cite{wang2020end}, so we extend the E2E-VC based DSR with the speaker encoder introduced in AC \cite{liu2020end} to preserve speaker identity.   

\section{Proposed method}
\begin{figure}[t]
  \centering
  \centerline{\includegraphics[width=8.5cm]{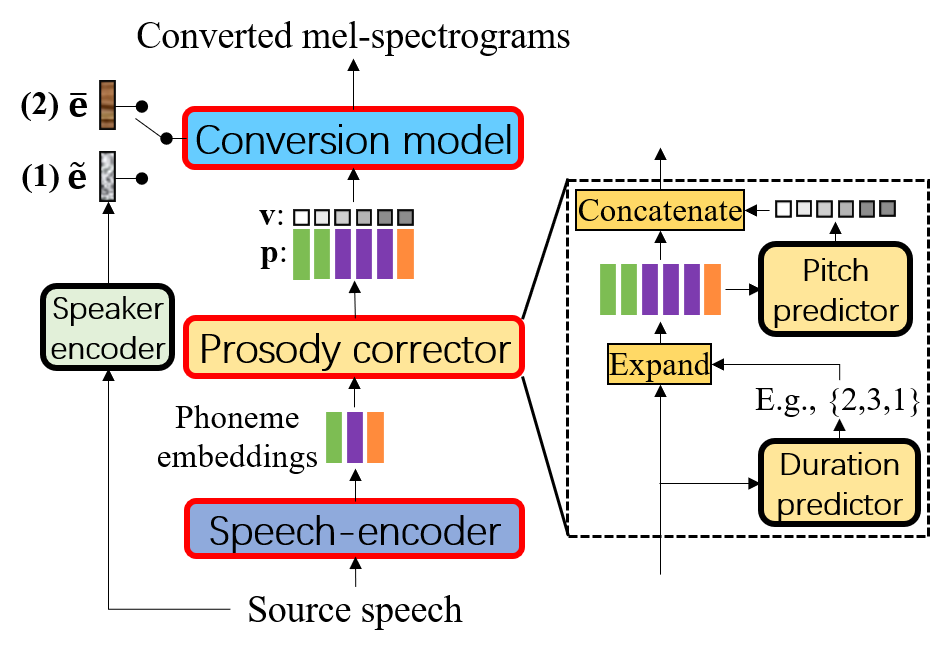}}
  \vspace{-1em}
  \caption{Diagram of the proposed VC system: (1) DSE $\widetilde{\textbf{e}}$ is extracted from the speaker encoder, which corresponds to Enc-CM; (2) DSE $\overline{\textbf{e}}$ is obtained by joint learning with the conversion model, which corresponds to Ada-CM.}\label{VC}
  \vspace{-2em}
\end{figure}

This section elaborates on the proposed VC approach with explicit prosodic modelling and DSE learning. The main differences from the baseline E2E-VC approach lie in two aspects: (1) Prosody is modelled in an explicit manner, so the prosody of the converted speech can be effectively controlled and corrected; (2) Speaker adaptation is proposed to obtain more effective DSE that is strongly related with speaker characteristics, leading to higher speaker similarity. As shown in Figure \ref{VC}, the whole VC system consists of three key modules, i.e., speech-encoder, prosody corrector and conversion model.

\vspace{-0.2em}
\subsection{Speech-encoder for phoneme embeddings extraction}
\label{ssec:speechEncoder}
\vspace{-0.05cm}
To preserve the linguistic content of original atypical speech, a speech-encoder is used to extract robust linguistic representations. Following \cite{wang2020end,liu2020end}, the speech-encoder adopts a seq2seq network to predict phoneme sequence. The speech-encoder is first pre-trained on large-scale typical speech data, then fine-tuned on the atypical speech of the dysarthric or L2 speaker $s_k$ to improve phoneme prediction accuracy. The pre-trained and fine-tuned speech-encoders are denoted as $\Phi_p$ and $\Phi _{{s_k}}$ respectively. We adopt the speech-encoder outputs that denote the phoneme probability distribution as the phoneme embeddings.

\vspace{-0.2em}
\subsection{Prosody corrector for explicit prosodic modelling}
\label{ssec:prosody}
\vspace{-0.05cm}
As atypical speech has atypical prosody, e.g., phoneme duration and pitch values, we propose explicit prosodic modelling by designing a prosody corrector to amend the atypical prosody to its typical version. As shown in Figure \ref{VC}, the prosody corrector contains the phoneme duration and pitch predictors, where the pitch can be described by fundamental frequency ($\textit{F}_0$). Both duration and $\textit{F}_0$ predictors are trained with L1 loss by using typical speech of a single speaker: (1) For duration prediction, the inputs are phoneme embeddings extracted by the speech-encoder $\Phi_p$ using the teacher-force mode. The targets are ground-truth phoneme durations, which are obtained from pairs of text and audio by forced alignment with Montreal Forced Aligner \cite{mcauliffe2017montreal}. (2) For $\textit{F}_0$ prediction, the expanded phoneme embeddings \textbf{p} by using the ground-truth phoneme duration are used as the inputs, and the targets are the ground-truth $\textit{F}_0$, denoted by \textbf{v}, which has the same length of \textbf{p}. When the duration and $\textit{F}_0$ predictors are well-trained, the prosody corrector is expected to infer typical phoneme duration and $\textit{F}_0$ values that are used to replace their abnormal counterparts for typical speech generation.

\vspace{-0.1cm}

\subsection{Conversion model for speech generation}
\label{ssec:conversion}
As shown in Figure \ref{VC}, we adopt a conversion model with function \textit{f} and parameters \textbf{W} to generate mel-spectrograms \textit{f}(\textbf{p},\textbf{v};\textbf{W},\textbf{e}), where the spoken content and duration are both controlled by the expanded phoneme embeddings \textbf{p}, the pitch and speaker identity are separately controlled by $\textit{F}_0$ \textbf{v} and DSE \textbf{e}. \textbf{e} is repeated and concatenated with \textbf{p} and \textbf{v} for generation. Given the typical speech of a set of speakers \textit{S} and atypical speech of a dysarthric or L2 speaker $s_k$, let $\textit{T}_{{s_i}}$ and $\textit{T}_{{s_k}}$ denote the set of mel-spectrogram features for speaker \(\textit{s}_i\) (\(\textit{s}_i\)$\sim$\textit{S}) and speaker $s_k$, respectively. Two DSE learning approaches are investigated and incorporated into the conversion model, i.e., speaker encoder based conversion model (Enc-CM) and speaker adaptation based conversion model (Ada-CM). For clarity, we denote the DSE used in Enc-CM and Ada-CM as $\widetilde{\textbf{e}}$ and $\overline{\textbf{e}}$ respectively.

\vspace{-0.5em}

\subsubsection{Speaker encoder based conversion model}
\label{sssec:Enc-CM}
The speaker encoder is a neural network for speaker verification and produces a fixed-dimensional DSE from acoustic feature frames of a speech utterance with variable length. It is trained to optimize a generalized end-to-end (GE2E) loss for DSE learning \cite{wan2018generalized}, so that the DSEs extracted from utterances of the same speaker and different speakers have high and low similarity, respectively. The DSE $\widetilde{\textbf{e}}_{{s_i}}$ derived from the speaker encoder is expected to capture speaker characteristics of $s_i$. By using typical speech data, Enc-CM is trained to minimize a loss \textit{L} (e.g., L1 loss) measuring the distance between the predicted and ground-truth mel-spectrograms:
\begin{footnotesize}
\begin{equation}
    \widetilde{\textbf{W}}_{SE}\!=\!\mathop{argmin}\limits_{\textbf{W}}\!{{E}_{{{s}_{i}}\sim{S},{{\textbf{a}}_{i,j}}\sim{{T}_{{{s}_{i}}}}}}\!\{\textit{L}(\textit{f}\:(\textbf{p}_{{i,j}},\!\textbf{v}_{{i,j}};\!\textbf{W},\!\widetilde{\textbf{e}}_{{s_i}}),\!\textbf{a}_{{i,j}})\}
\end{equation}
\end{footnotesize}where $\textbf{p}_{{i,j}}$ are phoneme embeddings extracted by speech-encoder $\Phi_p$ and expanded by ground-truth duration, $\textbf{v}_{{i,j}}$ and $\textbf{a}_{{i,j}}$ are ground-truth $\textit{F}_0$ and mel-spectrograms for speaker \(\textit{s}_i\) (\(\textit{s}_i\)$\sim$\textit{S}), respectively. 

At the conversion phase, the atypical speech of the speaker $s_k$ is used as the input of the speaker encoder and the fine-tuned speech-encoder $\Phi _{{s_k}}$ to extract DSE $\widetilde{\textbf{e}}_{{s_k}}$ and phoneme embeddings, respectively. The phoneme embeddings are used as the inputs of the prosody corrector to obtain the expanded phoneme embeddings $\hat{\textbf{p}}$ with typical duration and $\textit{F}_0$ $\hat{\textbf{v}}$. Finally, the system generates converted mel-spectrograms as \textit{f}($\hat{\textbf{p}}$,$\hat{\textbf{v}}$;$\widetilde{\textbf{W}}_{SE}$,$\widetilde{\textbf{e}}_{{s_k}}$). 

\subsubsection{Speaker adaptation based conversion model}
\label{sssec:Ada-CM}
Instead of obtaining the speaker representations from an external network, DSE can be jointly learned with the conversion model. The joint learning enables the DSE to directly capture the speaking characteristics related with speech generation, leading to higher speaker similarity. Specifically, Ada-CM involves two-stage training: First, the conversion model is pre-trained with typical speech data, where the DSE $\textbf{e}_{{s_i}}$ for each speaker ${s_i}$ is randomly initialized and jointly trained with \textbf{W}:  
\begin{footnotesize}
\begin{equation}
    \hat{\textbf{W}}_{SA}\!,\!\{\hat{\textbf{e}}_{{s_i}}\}\!=\!\mathop{argmin}\limits_{{\textbf{W},\{\textbf{e}}_{{s_i}}\}}\!{{E}_{{{s}_{i}}\sim{S},{{\textbf{a}}_{i,j}}\sim{{T}_{{{s}_{i}}}}}}\!\{\textit{L}(\textit{f}\:(\textbf{p}_{{i,j}},\!\textbf{v}_{{i,j}}\!;\!\textbf{W},\!\textbf{e}_{{s_i}}),\!\textbf{a}_{{i,j}})\}
\end{equation}
\end{footnotesize}Second, with the speech data of multiple speakers for training, the conversion model $\hat{\textbf{W}}_{SA}$ has good generalization capacity and can be fine-tuned well to unseen speakers for DSE learning. Therefore, for the dysarthric or L2 speaker $\textit{s}_k$ with the expanded phoneme embeddings $\textbf{p}_{{k,j}}$ and $\textit{F}_0$ $\textbf{v}_{{k,j}}$, speaker adaptation is performed as: 
\begin{footnotesize}
\begin{equation}
    \overline{\textbf{W}}_{SA},\!\overline{\textbf{e}}_{{s_k}}\!=\!\mathop{argmin}\limits_{{\textbf{W},\textbf{e}}_{{s_k}}}\!{{E}_{{{\textbf{a}}_{k,j}}\sim{{T}_{{{s}_{k}}}}}}\!\{\textit{L}(\textit{f}\:(\textbf{p}_{{k,j}},\!\textbf{v}_{{k,j}}\!;\!\textbf{W},\!\textbf{e}_{{s_k}}),\!\textbf{a}_{{k,j}})\}
\end{equation}
\end{footnotesize}where \textbf{W} is initialized by $\hat{\textbf{W}}_{SA}$ and DSE $\textbf{e}_{{s_k}}$ is also randomly initialized. After adaptation, target speaker characteristics that are beneficial for speech generation are encoded into $\overline{\textbf{e}}_{{s_k}}$. 

Similar with Enc-CM, at the inference phase, we can use the adapted conversion model $\overline{\textbf{W}}_{SA}$ to generate the converted mel-spectrograms as \textit{f}($\hat{\textbf{p}}$,$\hat{\textbf{v}}$;$\overline{\textbf{W}}_{SA}$,$\overline{\textbf{e}}_{{s_k}}$) with high speaker similarity achieved by $\overline{\textbf{e}}_{{s_k}}$, and typical prosody controlled by predicted typical phoneme duration and $\textit{F}_0$. 

\section{Experiments}
\label{sec:exp}

\subsection{Experimental settings}
\label{ssec:exp-settings}

Experiments are conducted on the LibriSpeech \cite{panayotov2015librispeech}, VCTK \cite{veaux2016superseded}, LJSpeech \cite{ito2017lj}, UASpeech \cite{kim2008dysarthric} and L2-ARCTIC \cite{zhao2018l2} datasets. Parallel WaveGAN (PWG) \cite{yamamoto2020parallel} is adopted as the vocoder to synthesize the waveform from the converted mel-spectrograms. We use 960h training data of LibriSpeech for pre-training the speech-encoder $\Phi_p$, 105 native speakers of VCTK for the training of PWG, and the training of Enc-CM and Ada-CM to obtain $\widetilde{\textbf{W}}_{SE}$ and $\hat{\textbf{W}}_{SA}$, respectively. The typical speech of single female speaker from LJSpeech is used for training duration and $\textit{F}_0$ predictors. For atypical speech, we select speaker M05 of UASpeech and speaker LXC of L2-ARCTIC for experiments. M05 has moderate-severe dysarthria with the speech having middle intelligibility. Following \cite{wang2020end}, we use the speech of blocks 1 and 3 for speech-encoder and Ada-CM fine-tuning, and the speech of block 2 for testing. As the audio of M05 has strong background noise which degrades the speaker adaptation performance, we adopt log-MMSE speech enhancement algorithm \cite{ephraim1985speech} to pre-process the audio. LXC speaker is a non-native English speaker with Mandarin accent and has 1131 recorded utterances, which are randomly divided into 1000/66/65 for training/validation/testing, where training and validation data are used for fine-tuning the speech-encoder and Ada-CM. The speech is sampled or resampled to 16kHz, and all speech features are calculated with 25ms Hanning window, 10ms frame shift and 400-point fast Fourier transform.

The speech-encoder has a similar architecture as in \cite{wang2020end}, including a 6-layer VGG extractor and a 5-layer BLSTM with 512 units per direction in the encoder, 512-dimensional location-aware attention and 2-layer LSTM with 1024 units in the decoder. The inputs to the speech-encoder are 40-band mel-spectrograms appended with delta and delta-delta features. Adadelta optimizer \cite{zeiler2012adadelta} with learning rate of 1 and batch size 8 is applied for the pre-training and fine-tuning of the speech-encoder with 1M and 2k steps, respectively. Duration and $\textit{F}_0$ predictors adopt the same structure, which consists of a 3-layer BGRU with 256 units per direction, 3 convolution layers with kernel size of 5, 9 and 19 respectively, and a 1-dimensional fully-connected (FC) layer to predict the duration or $\textit{F}_0$ value. Both predictors are trained by the Adam optimizer \cite{kingma2014adam} with a learning rate of 0.001, batch size of 16 and 30k steps. The settings and training of speaker encoder is same as \cite{liu2020end}, and DSE is set to 256-dimensional for both Enc-CM and Ada-CM. The conversion model is a frame-to-frame network composed of two 512-dimensional FC layers, 4-layer BLSTM with 512 units per direction and one 80-dimensional FC layer to predict mel-spectrograms. Both Enc-CM training and Ada-CM pre-training use the learning rate of 0.001 and batch size of 16 with 50k steps, and Ada-CM fine-tuning takes 3k steps. Readers are encouraged to listen to our audio samples\footnote{Audio samples: \url{https://wendison.github.io/VC-DSR-AC-demo/}}.

% More details can be found on our demo\footnote{Audio samples: \url{https://wendison.github.io/VC-DSR-AC-demo/}} and the code\footnote{Code: \url{https://github.com/Wendison/Atypical-voice-conversion}} that will be released soon.

We compare the Enc-CM and Ada-CM with our previously proposed E2E-VC for DSR \cite{wang2020end} and AC \cite{liu2020end}, where the original settings are adopted. To evaluate the performance of all methods, 20 listeners are invited to give subjective evaluations, including mean opinion score (MOS) tests (1-bad, 2-poor, 3-fair, 4-good, 5-excellent) to evaluate speech naturalness and speaker similarity, AB preference tests to evaluate the impact of phoneme duration and $\textit{F}_0$, and objective evaluation of speech intelligibility based on a speech recognition model. 
% Readers are encouraged to listen to generated audio samples\footnote{Audio samples: \url{https://wendison.github.io/VC-DSR-AC-demo/}}.

\subsection{Experimental results}
\label{ssec:results}

\subsubsection{Speech naturalness and speaker similarity comparison}
\label{sssec:MOS}
\begin{figure}[t]

\begin{minipage}[b]{1.0\linewidth}
  \centering
  \centerline{\includegraphics[width=6cm]{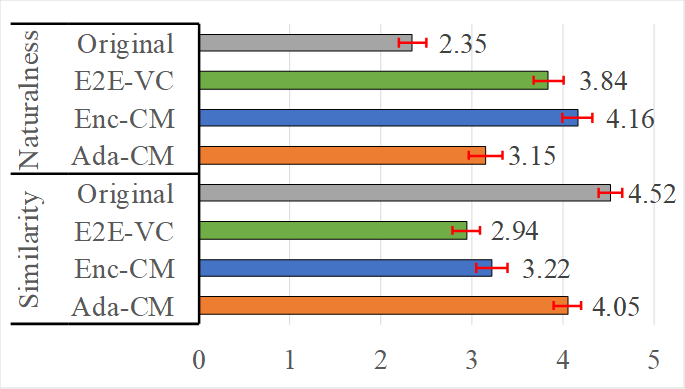}}
%  \vspace{2.0cm}
  \centerline{(a) DSR experiments}\medskip
\end{minipage}
%
% \vspace{-1.5em}
\begin{minipage}[b]{1.0\linewidth}
  \centering
  \centerline{\includegraphics[width=6cm]{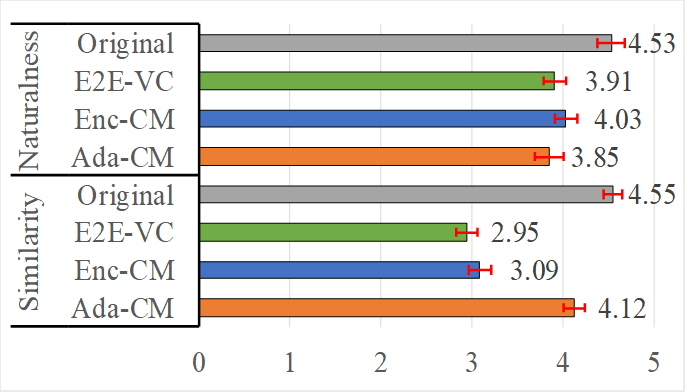}}
%  \vspace{2.0cm}
  \centerline{(b) AC experiments}\medskip
\end{minipage}
\vspace{-2.5em}
\caption{Comparison results of MOS with 95\% confidence intervals for speech naturalness and speaker similarity}
\label{MOS}
\vspace{-2em}
\end{figure}

Figure \ref{MOS} shows the MOS results for speech naturalness and speaker similarity, where ‘Original’ denotes the original dysarthric or L2 speech. We randomly select 15 testing utterances of M05 or LXC for evaluation. For DSR experiments as shown in Figure \ref{MOS}(a), we observe that compared with the original dysarthric speech, the converted speech by all methods achieves improvements in naturalness. The proposed Enc-CM achieves higher naturalness than E2E-VC, indicating that the effectiveness of the proposed prosody corrector to generate the speech with stable and accurate prosody. Ada-CM achieves lower naturalness than E2E-VC, partially due to the speech enhancement algorithm degrading the quality of M05 audio for speaker adaptation. Besides, speaker adaptation also inevitably incorporates the abnormal speaking characteristics of the dysarthric speaker into the converted speech, such as atypical prosody and articulation, which partially contribute to the speaker characteristics. As a result, Ada-CM can achieve highest speaker similarity, followed by Enc-CM and E2E-VC. For AC experiments as shown in Figure \ref{MOS}(b), all methods achieve high and similar speech naturalness, while Ada-CM also achieves the highest speaker similarity. This verifies the effectiveness of explicit prosodic modelling and DSE learned by speaker adaptation for achieving high speaker similarity. 

\vspace{-0.5em}
\subsubsection{Impact of phoneme duration and $\textit{F}_0$}
\label{sssec:Ablation}
\begin{figure}[t]
\begin{minipage}[b]{1.0\linewidth}
  \centering
  \centerline{\includegraphics[width=8cm]{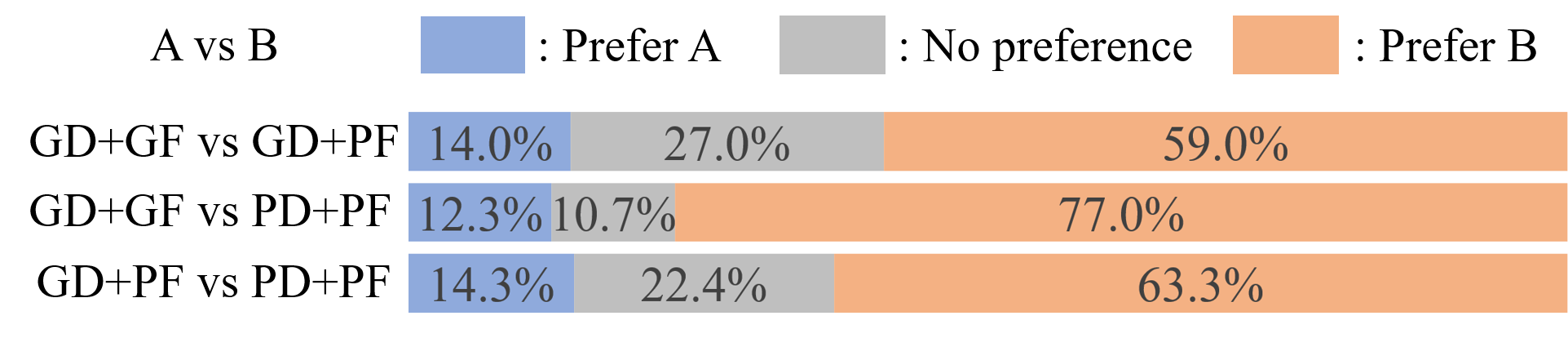}}
%  \vspace{-0.2cm}
  \centerline{(a) DSR experiments}\medskip
\end{minipage}
\begin{minipage}[b]{1.0\linewidth}
  \centering
  \centerline{\includegraphics[width=8cm]{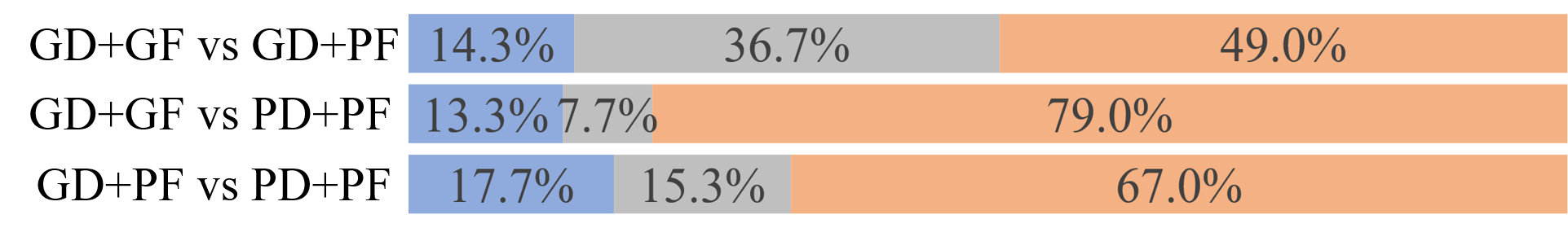}}
%  \vspace{-0.8cm}
  \centerline{(b) AC experiments}\medskip
\end{minipage}
\vspace{-0.9cm}
\caption{AB preference test results for different combinations
of phoneme duration and $\textit{F}_0$.}
\label{ablation}
\vspace{-1.em}
\end{figure}

We investigate the impact of the phoneme duration and $\textit{F}_0$ on the normality and accentedness of the converted speech. We explore three combinations of phoneme duration and $\textit{F}_0$ used by Enc-CM to generate the speech: (1) Ground-truth Duration and Ground-truth $\textit{F}_0$ (GD+GF); (2) Ground-truth Duration and Predicted $\textit{F}_0$ (GD+PF); (3) Predicted Duration and Predicted $\textit{F}_0$ (PD+PF). The AB preference test is conducted, and results are shown in Figure \ref{ablation}. From the comparison ‘GD+GF vs GD+PF’, we can see that the predicted typical $\textit{F}_0$ facilitates the conversion model to generate near-normal or near-native speech. From the comparison ‘GD+PF vs PD+PF’, we observe that using the predicted typical duration can further improve the quality of the converted speech, this shows that both phoneme duration and $\textit{F}_0$ affect speech normality or the degree of accentedness, and the proposed prosody corrector is helpful for attaining typical phoneme duration and $\textit{F}_0$ values, which are beneficial for generating speech with normal or native prosody characteristics.

\vspace{-0.5em}
\subsubsection{Speech intelligibility comparison}
To show the effectiveness of proposed methods to improve the intelligibility of atypical speech, a publicly released automatic speech recognition model, i.e., Jasper \cite{li2019jasper}, is used to test the character error rate (CER) and word error rate (WER) with greedy decoding. The results are illustrated in Table \ref{asr}, we also report the results for 'Original (Mel+PWG)' that uses the original mel-spectrograms to synthesize the waveform by using PWG vocoder. We can see that 'Original (Mel+PWG)' is inferior to 'Original', which indicates that the PWG vocoder tends to degrade the speech quality. For DSR experiments, we can observe that CER and WER of the original dysarthric speech can be significantly reduced by the proposed methods, where Enc-CM performs the best and achieves 47.6\% and 29.3\% absolute reduction for CER and WER respectively. As the original dysarthric speech used in Ada-CM to perform speaker adaptation contains strong background noise, even though log-MMSE is used for denoising, the pre-processed audio still contains artificial noise that hurts Ada-CM performance, thus smaller CER and WER reduction is achieved for Ada-CM. For AC experiments, the proposed Enc-CM can still achieve 2.4\% CER and 4.4\% WER reduction compared with Original speech, the baseline E2E-VC and proposed Ada-CM have no improvements over Original speech while achieve similar CER and WER with 'Original (Mel+PWG)', adopting a more powerful vocoder is expected to enhance the speech intelligibility.

% \vspace{-2em}

\begin{table}[t]
  \caption{Comparisons based on CER (\%) and WER (\%).}
  \label{asr}
  \vspace{-1em}
  \centering
  \scalebox{0.96}{
  \begin{tabular}{c|c|c|c|c}
    \thickhline
    \multirow{2}{*}{Methods} & \multicolumn{2}{|c}{DSR experiments} & \multicolumn{2}{|c}{AC experiments} \\
    \cline{2-5}
    & CER & WER & CER & WER
    \\
    \hline
    Original & 90.2 & 91.0 & 17.7 & 35.5 \\
    Original (Mel+PWG) & 94.3 & 95.3 & 22.4 & 40.8 \\
    \hline
    E2E-VC & 50.6 & 69.8 & 22.7 & 41.3 \\
    Enc-CM & \textbf{42.6} & \textbf{61.7} & \textbf{15.3} & \textbf{31.1} \\
    Ada-CM & 56.5 & 80.5 & 21.5 & 40.2 \\
    \thickhline
  \end{tabular}}
  \vspace{-2em}
\end{table}

% \vspace{-2cm}

\section{Conclusions}
\label{sec:con}
This paper presents a VC system for converting atypical speech to typical speech, by explicit prosodic modelling and DSE learning. prosodic modelling is proposed to leverage phoneme duration and $\textit{F}_0$ predictors to obtain typical values for prosody correction, while speaker encoder and speaker adaptation approaches are separately proposed to obtain effective DSE to capture speaker characteristics. DSR and AC experiments show that proposed methods can achieve reduction of dysarthric and non-native speaking characteristics, where significant intelligibility improvements can be achieved for dysarthric speech. Enc-CM outperforms previously proposed E2E-VC, and Ada-CM achieves the highest speaker similarity. However, for Ada-CM, atypical pronunciation patterns are also incorporated into the converted speech after speaker adaptation. Explicitly modelling more para-linguistic information may be helpful to mitigate this problem. In addition, using better speech denoising algorithms or cleaner audio data is expected to further improve the performance of Ada-CM, this will be studied in the future. 

\section{Acknowledgements}
This research is partially supported by a grant from the HKSARG Research Grants Council General Research Fund (Project Reference No. 14208817).


\end{document}